# De-Pinning Transition of Bubble Phases in a High Landau Level

Xuebin Wang[1], Hailong Fu[1], Lingjie Du[2], Xiaoxue Liu[1], Pengjie Wang[1], L. N. Pfeiffer[3], K. W. West[3], Rui-Rui Du*[,1,2,4], Xi Lin*[,1,4]

[1] International Center for Quantum Materials, Peking University, Beijing 100871, China
[2] Department of Physics and Astronomy, Rice University, Houston, Texas 77251-1892, USA
[3] Department of Electrical Engineering, Princeton University, Princeton, New Jersey 08544, USA
[4] Collaborative Innovation Center of Quantum Matter, Beijing 100871, China

While in the lowest Landau level the electron-electron interaction leads to the formation of the Wigner crystal, in higher Landau levels a solid phase with multiple electrons in a lattice site of crystal was predicted, which was called the bubble phase. Reentrant integer quantum Hall states are believed to be the insulating bubble phase pinned by disorder. We carry out nonlinear transport measurements on the reentrant states and study the de-pinning of the bubble phase, which is complementary to previous microwave measurements and provides unique information. In this study, conductivity is directly measured with Corbino geometry. Based on the threshold electric field of de-pinning, a phase diagram of the reentrant state is mapped. We discuss an interaction-driven topological phase transition between the integer quantum Hall state and the reentrant integer quantum Hall state.

A two-dimensional electron gas (2DEG) displays the integer quantum Hall effect (IQHE) and the fractional quantum Hall effect (FQHE) when it is subject to a perpendicular magnetic field B and an ultra-low temperature environment. In the high magnetic field limit, if the filling factor ($\upsilon = \frac{n}{eB/h}$, where $n$ is the electron density) is small, electrons tend to form into an electron solid state rather than the fractional quantum Hall states, known as the Wigner crystal (WC) [1,2]. At a modest magnetic field, many-body phases different from the fractional quantum Hall states can be more favorable in high Landau levels (LLs) [3-8], such as the stripe state and the bubble state. Filling factor 9/2 is well known as a candidate for the stripe state [9,10] and the reentrant integer quantum Hall effect (RIQHE) is believed to be the candidate for the bubble state [3,11,12], analogous to the WC but with multiple electrons per site.

RIQHE has been observed at partial filling factors between integer quantum Hall states or between fractional quantum Hall states [9,10,13-18]. Being pinned by disorder, electrons in the partially filled LLs do not contribute to RIQHE's regular transport properties. In a square or Hall bar geometry, RIQHE is similar to IQHE with vanishing longitudinal resistance and with integer quantization of Hall resistance reentrant to the neighboring quantum Hall plateau. In the Corbino geometry, both RIQHE and IQHE exhibit zero conductivity. In Corbino samples the edge states are shunted by electrodes [19] and therefore the transport is free of Hall contributions. Specifically, with Corbino samples the conductivity is directly measured and the electric field across the samples can be precisely controlled.

RIQHE between ν=4 and ν=5 has been observed in the square geometry and the nonlinear transport measurement provides evidence for the insulating bubble state [20,21], with similar techniques used in probing the WC [22-24]. Their filling factors are generally believed to be ~4.25 and ~4.75 [9,10,16,18]. Based on Hartree-Fock calculations, the ground state of 2DEG in partially filled high LLs is proposed to be a charge density wave (CDW), and the partial filling factor determines whether the ground state is a bubble phase as well as the number of electrons per lattice site [3,11,12,25-27]. Microwave experiments have provided evidence for the bubble phase through a resonance in the real part of the diagonal conductivity [28]. Same features have been found in microwave experiments on the WC at filling factors ν<1/5 [23].

At high magnetic fields, quantum fluctuations in real space (approximately equal to the magnetic length $l$, $l = \sqrt{\hbar/eB}$) are in the order of the lattice constant so only the WC survives in the dilute density limit. However, in high LLs for $N>0$, and typically with moderate magnetic fields, the Hartree-Fock ground state becomes relevant. Due to the much larger lattice constant (in the order of $3.3\sqrt{2N+1} \cdot l$), fluctuations do not induce the melting of long-range crystalline order even if the partially filled electrons or holes are still far away from the dilute limit. An obvious alternative of CDW in high LLs is FHQE and numerical results are in favor of CDW [6,11,12,25-27,29,30]. The one-, two- and three-electron bubble phases and stripe phase are predicted in the third LL by the Hartree-Fock theory but three-electron bubbles are not expected in the density matrix renormalization group (DMRG) calculations [30]. The transition between bubble states and the WC in high LLs is discussed as a first order phase transition [30]. The melting temperature of CDW is also estimated [11]. In the zero-temperature limit, when quantum fluctuations are taken into account, new phases such as smectic and nematic states should emerge [5,8].

In this Letter, we present findings in the reentrant states between $\nu=4$ and $\nu=5$ in the Corbino geometry [19]. There exists a threshold electric field, beyond which the differential conductivity deviates from zero and grows rapidly. This kind of nonlinear transport implies that the insulating nature of the reentrant states is a bubble phase pinned by the impurity. At the lowest temperature, reentrant states merge with the nearby integer states and nonlinear transport provides a useful tool to distinguish the bubble phase, an interaction induced insulating state, from the integer quantum Hall state, a single particle insulating state. Based on these results we have arrived at a schematic phase diagram of reentrant states as a function of temperature and filling factor.

The measurements were made on two GaAs/AlGaAs heterostructures. The densities of sample A and sample B are $4.2\times10^{11}$ cm$^{-2}$ and $2.8\times10^{11}$ cm$^{-2}$, and their motilities are $2.1\times10^{7}$ cm V$^{-1}$ s$^{-1}$ and $2.8\times10^{7}$ cm V$^{-1}$ s$^{-1}$. Samples were illuminated by a red LED at 4.5 K and 15 μA for one hour. The measurements were carried out in a dilution fridge with a base temperature of 6 mK (measured at the mixing chamber) and a base electron temperature of about 25 mK. Longitudinal and Hall resistance signature of RIQHE is confirmed by a square sample from the same wafer (Fig. 1a). Corbino samples have a ring of 2DEG with 100 μm in width. The two terminal differential conductivity measurements were performed using the standard lock-in techniques at 23 Hz, as shown in Fig. 1b and Fig. 1c. Maximum AC voltage drop across Corbino samples was 50 μV.

In Fig. 1a, a square sample clearly shows the anisotropic transport at $\nu=9/2$ and the signature of reentrant states at ~4.25 and ~4.75 filling factors. In Fig. 1b, Corbino sample A shows insulating behavior for both IQHE and RIQHE at a mixing chamber temperature of 100 mK, and the RIQHE=4 reentrant insulating state is separated from its neighboring IQHE=4 insulating state by a conducting region in magnetic fields. At lower temperatures, the conducting region vanishes, and two insulating states merge with each other. Regardless of the merging, RIQHE at the base temperature can still be identified either from the comparison with a square sample or with conductivity data from 100 mK. The filling factors and energy gaps for RIQHE=4 and RIQHE=5 are 4.25, 4.70, 1.56 K and 1.24 K respectively. In RIQHE, if electric fields are applied to the Corbino samples by the DC bias, the de-pinning transition from the insulating state appears. The inset of Fig. 1b shows an example of nonlinear behaviors in

RIQHE, similar to what have been measured in the square sample's RIQHE [20] and in the WC [23]. The rapid disappearance of the insulating state, the irregular differential conductivity above the threshold electric field, and the hysteresis are evidence of the de-pinning of the bubble phase. In this work, the threshold electric field, $E_T$, is defined as the critical value below which the differential conductivity remains zero based on our experimental resolution, taken from a sweep from zero to positive electric fields. The stronger a bubble phase is, the larger the difference of the threshold electric fields between sweep-up and sweep-down is, and the stronger the hysteresis is. Similar sliding behaviors in RIQHE are also observed at other temperatures and magnetic fields for both samples.

The conductivity of Corbino sample B at different temperatures is shown in Fig. 2a. At the lowest temperature, RIQHE=5 merges with IQHE=5, which is different from sample A. The filling factors and energy gaps for RIQHE=4 and RIQHE=5 are 4.41, 4.67, 0.83 K and 3.01 K respectively (Fig. 2b). Inset of Fig. 2a shows the energy gap as a function of the magnetic field. The energy gap between RIQHE and IQHE does not close. Fig. 2c shows the threshold electric field as a function of the magnetic field at the lowest temperature, where the RIQHE=5 insulating state fully develops and merges with IQHE=5. Note that another phenomenon called breakdown will also destroy IQHE's insulating state [31]. Fig. 2d shows examples of nonlinear transport in IQHE, transition from IQHE to RIQHE, and RIQHE. With similar threshold electric fields, the difference between breakdown and de-pinning is significant. The conductivity in breakdown changes slowly and smoothly above the threshold electric field. Although the conductivity is zero from $\nu$=4.6 to $\nu$=5.0, the nature of insulating can result from either IQHE or RIQHE, and nonlinear transport presents evidence for both states.

Nonlinear responses of longitudinal conductivity to electric fields are systematically measured in sample B at different temperatures and magnetic fields. The phase diagram of RIQHE, which is mapped according to the threshold electric field, clearly shows that the RIQHE=5 and IQHE=5 are well separated (Fig. 3a). Interestingly, the most stable filling factor in terms of the melting temperature is not the most stable point for the electric field (Fig. 3b). The strongest reentrant effects for the temperature and the electric field occur at different filling factors, which excludes the possibility of the thermal heating breakdown process for the nonlinear behavior of RIQHE. In this work, the filling factors of RIQHE are determined by the position of the minimum conductivity when the temperature is high enough to melt the bubble phases. Theoretical calculations suggest that the number of electrons per bubble depends on the partial filling factor [25,26,30]. Filling factor of 4.41 for RIQHE implies bubbles of three electrons, different from two-electron bubbles at ~4.25. A recent experimental study [18] points out that the onset temperatures of RIQHE in the second and third LLs are inconsistent with calculated cohesive energy [26,32], and suggests that the number of electrons per bubble in the third LL is likely different from what was previously predicted. In reference [18], the ratio of reduced onset temperature between RIQHE=4 and RIQHE=5 is slightly larger than 1, which is similar to our results in sample A. The reduced onset temperature is defined as $k_B T_c/E_c$, where $T_c$ is the onset temperature determined by a sharp peak in longitudinal resistance [16] and $E_c$ is the Coulomb energy. In sample B, the reduced onset temperature is $11.2 \times 10^{-4}$ for RIQHE=4 and $57.4 \times 10^{-4}$ for RIQHE=5, where we use the disappearance of zero conductivity to determine the onset temperature. The reduced onset temperature provides a contrast for energy scales of different reentrant states. Judging from the different melting temperatures of RIQHE between sample A and sample B, it is likely that the number of electrons per bubble depends on factors more than the interaction between electrons and the magnetic field. Another recent experimental work in the second LL also suggests that bubble phases are more complicated than currently anticipated [33].

The lattice constant of the bubble crystal is approximately $3.3R_c$, where cyclotron radius $R_c = \sqrt{2N+1} \cdot l$, $l$ is the magnetic length and $N$ is the LL index [11]. Given the wafers and the reentrant states we study, the lattice constant is around 100 nm. Bubbles form domains that slide under the electric field. We can make a simple assumption that the domain pinning potential is comparable to the thermal activated energy plus the external electric energy, which is the charge under the electric field over the domain size. Therefore, $A \times e \times d \times E_T \sim -k_B T$, where $A$ is the number of electrons per site and $d$ is the domain characteristic size. The observed threshold electric field shows a nearly negative linear dependence on the temperature in this work, indicating a domain size order of 0.1 to 1 μm at different reentrant states. Domain size can also be estimated through theoretical work [34] where the relationship between the threshold electric field and the linear dimension of decorrelated domains is provided. Based on Equation (5.1) in [34], we get the same order of magnitude in domain sizes. The ratios of domain size over lattice constant are ~13 (Sample A, RIQHE=4), ~16 (Sample A, RIQHE=5), ~20 (Sample B, RIQHE=4), ~3 (Sample B, RIQHE=5) at the center of reentrant and at the lowest temperature.

The merging of RIQHE and IQHE is an interesting observation, which happens in both Corbino samples. The merging of zero longitudinal resistance and quantized Hall resistance is also observed at the lowest temperature in the square sample. The transition from RIQHE to IQHE is a Chern number=4 (or 5) topological state to another topological state with the same Chern number, and it should go through a metallic state. At a higher temperature, RIQHE and IQHE are both zero conductivity states, and they are separated by a conductivity peak, indicating an insulator-metal-insulator transition as expected. Our measurement does not show such a metallic state at the lowest temperature and the two phases are indistinguishable in transport (schematically depicted in Fig. 3c). We believe that such observation can be described by a quantum phase transition. Mixed insulating phases between the bubble phase and the WC have been proposed by both microwave and transport measurements [35,36]. Inhomogeneity of the 2DEG, and the separation between

IQHE regions and patches of solids may contribute to the phenomenon we observed. Similar mixing of two different phases has been suggested near filling factor 1/3 state [37].

In conclusion, we report transport measurements of the insulating states between $\nu=4$ and $\nu=5$. We have demonstrated a direct measurement of conductivity in the reentrant states, and established the de-pinning transition of the bubble phases. The nonlinear transport results of IQHE and RIQHE have shown distinct characteristics relevant to breakdown and sliding, respectively. The sliding of the reentrant state is studied at various temperatures and magnetic fields, and the phase diagram of RIQHE is provided. These observation leads to the conclusion of de-pinning of the bubble phases. At the lowest temperature, IQHE merges with RIQHE, indicating a transition from a single particle insulating state to a many body insulating state with the same Chern number.

Acknowledgement:

We are grateful to Xincheng Xie, Fa Wang, Chi Zhang and Junren Shi for their helpful comments. The work at PKU was funded by NBRPC (Grant No. 2012CB921301) and NSFC (Grant No. 11274020 and 11322435). The work at Rice was funded by DOE (DE-FG02-06ER46274). The work at Princeton was partially funded by the Gordon and Betty Moore Foundation as well as the National Science Foundation MRSEC Program through the Princeton Center for Complex Materials (DMR-0819860).

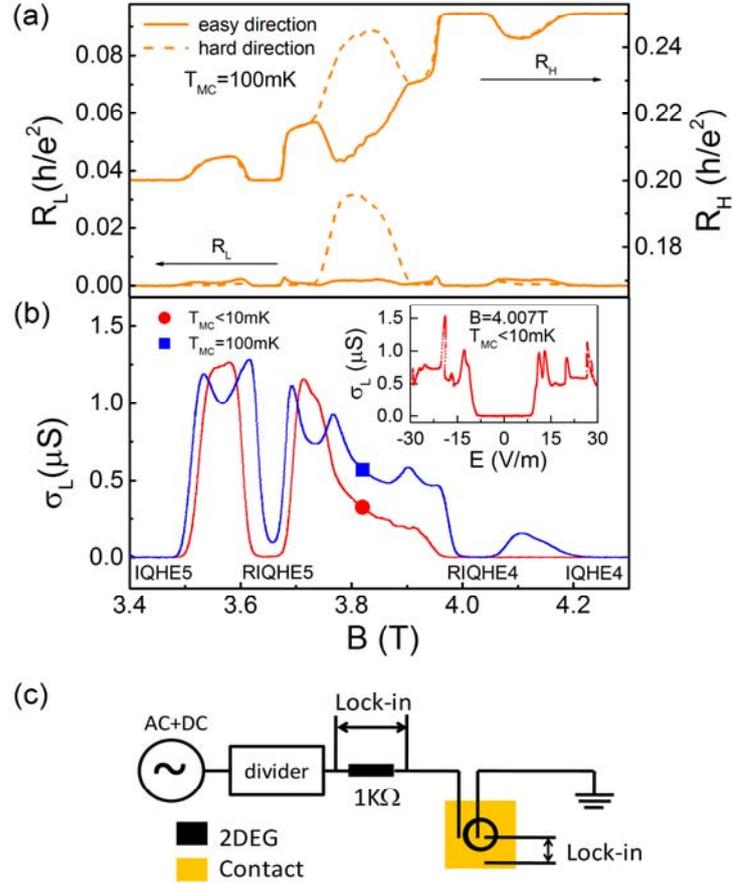

FIG. 1. (a) Longitudinal and Hall resistance of a square sample made from the same wafer as sample A. The signature of RIQHE is clearly identified. (b) Longitudinal conductivity of Corbino sample A at two temperatures, normalized by the sample geometry. RIQHE=4 is stronger than RIQHE=5. RIQHE=4 merges with IQHE=4 at the lowest temperature. Inset: an example of sliding for RIQHE. (c) Differential conductivity measurement setup for Corbino samples. The electric field across the sample is provided by the DC voltage drop between inner and outer contacts divided by the 2DEG ring width.

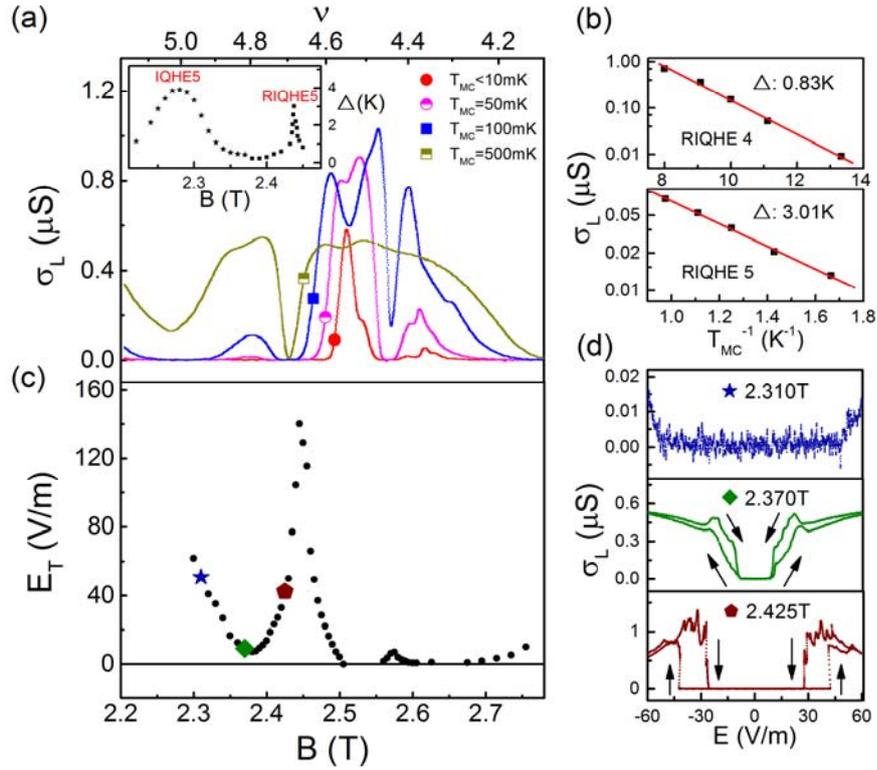

FIG. 2. (a) Differential conductivity of Corbino sample B as a function of the magnetic field at different temperatures. RIQHE=5 is stronger than RIQHE=4. RIQHE=5 merges with IQHE=5 at the lowest temperature. Inset is the energy gap $\Delta$ as a function of magnetic field, showing a transition from IQHE to RIQHE. Below 2.375 T (symbol star), the energy gap is fitted by the equation $\sigma_L \propto \exp(-\Delta/2k_BT)$. Above 2.375 T (symbol square), the energy gap is fitted by the equation $\sigma_L \propto \exp(-\Delta/k_BT)$. (b) Examples of energy gap of reentrant states, fitted by the equation $\sigma_L \propto \exp(-\Delta/k_BT)$. (c) Threshold electric field $E_T$ as a function of magnetic field from IQHE=5 to RIQHE=5 at the lowest temperature. (d) From top to bottom, they are examples for IQHE breakdown, crossover behavior from IQHE=5 to RIQHE=5, and RIQHE de-pinning, respectively. The examples at three filling factors are labeled differently in (c).

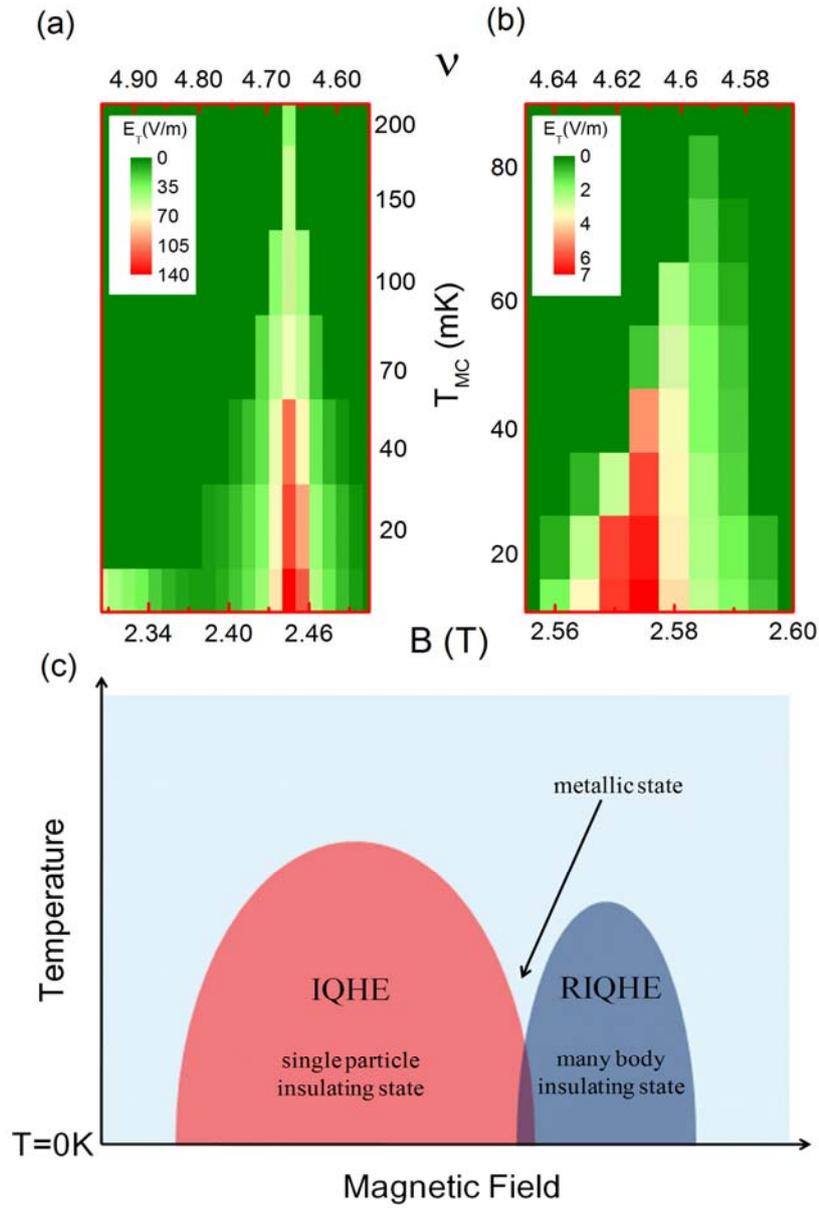

FIG. 3. (a) B-T phase diagram of RIQHE=5 for sample B, different colors represent different values of threshold electric fields. (b) B-T phase diagram of RIQHE=4 for sample B, different colors represent different values of threshold electric fields. (c) Schematic view of phase diagram for the transition from IQHE=5 to RIQHE=5.